# Generalized Adaptive Network Coded Cooperation (GANCC): A Unified Framework for Network Coding and Channel Coding

Xingkai Bao, *Student member, IEEE* and Jing Li (Tiffany), *Senior Member, IEEE*

Department of Electrical and Computer Engineering,

Lehigh University, Bethlehem, PA 18015

Email: { xib3, jingli}@ece.lehigh.edu

## Abstract

This paper considers distributed coding for multi-source single-sink data collection wireless networks. A unified framework for network coding and channel coding, termed *generalized adaptive network coded cooperation* (GANCC), is proposed. Key ingredients of GANCC include: matching code graphs with the dynamic network graphs on-the-fly, and integrating channel coding with network coding through circulant low-density parity-check codes. Several code constructing methods and several families of sparse-graph codes are proposed, and information theoretical analysis is performed. It is shown that GANCC is simple to operate, adaptive in real time, distributed in nature, and capable of providing remarkable coding gains even with a very limited number of cooperating users.

## Index Terms

User cooperation, network coding, channel coding, low-density parity-check (LDPC) codes, density evolution

## I. INTRODUCTION

Inherent to networked communication is the cooperation among different users which pulls together all dimensions of resources[1]-[4]. User cooperation may occur in different forms, among different numbers of users, and at different layers of the network. In the physical layer, user cooperation rooted back to the relay channel problem in the seventies, and has experienced an amazing comeback the last few years. In the network layer, user cooperation has taken the form of cooperated routing and resource management. User cooperation is particularly beneficial

This work is supported in part by National Science Foundation under Grant No. CCF-0430634 and CCF-0635199, and by the Commonwealth of Pennsylvania, Department of Community and Economic Development, through the Pennsylvania Infrastructure Technology Alliance (PITA).

for wireless systems – while an individual channel operating alone may be useless due to severe path loss, shadowing or fading, combined together a set of channels may become useful again.

This paper studies cooperative strategies for wireless ad-hoc networks that comprise a host of users communicating with a common destination. Leveraging the technologies from the physical and the network layer, we consider the joint treatment of channel coding and network coding to combat fading or random link outage. Whereas channel coding has long been the established technology for protecting bit streams, network coding has only recently found its way here. Originated from the network flow problem, network coding generalizes traditional store-and-forward routing, by allowing intermediate relaying nodes to perform simple coding operation [5]-[6]. Network coding allows for a higher throughput than traditional routing, and, in the wireless context, also increases the diversity order and reduce the outage probability [8]-[12].

Existing network coding schemes (e.g. [11], [12]) are predominantly based on pre-determined fixed network codes, and may therefore fall miserably to the random link outage caused by fading, shadowing and other unforeseen wireless conditions. In contrast, the *adaptive network coded cooperation* (ANCC) protocol developed in [8], [10] adaptively generates network codes – by matching, in real time, the code graph of some low density parity check (LDPC) code with the network graph that specifies the instantaneous network topology. It thus provides a practical and efficient solution to the variant and unstable nature of wireless links and network topology, a concern that had previously limited the exploitation of network coding in the wireless scenario.

The ANCC protocol assumes that channel coding is performed separately from routing at the edge of the network. Although seemingly convenient, this constraint is unnecessary as well as suboptimal. This paper generalizes ANCC by unifying channel coding and network coding. The idea finds its motivation and theoretic support in the emerging network information theory. Well-known from the Shannon information theory is the source-channel separation for *channels*, which states that source coding and channel coding can be performed independently over a communication channel without losing optimality. Recent studies indicate that source-channel separation may also hold for *networks*, but source-network separation and channel-network separation will break [7]. Hence, while source coding and channel coding may still be treated separately in such network scenarios as multiple access and broadcast channels, separating routing (network coding) from source or channel coding will lose end-to-end optimality.

In a separate channel-network coding treatment, although it is possible to perform iterative channel-network decoding to extract as much gain as possible, rate allocation between the network code and the channel code becomes a sensitive issue, whose improper design may detrimentally affect the performance. Rather than simply concatenating channel codes and net-work codes, we treat channel coding as an integral part of network coding, thereby reducing this issue to a minimal. Following the notion developed in [7] that network codes are essentially generalization of source codes and channel codes, we refer to the new protocol as the *generalized adaptive network coded cooperation* (GANCC) protocol. GANCC makes clever use of circulant sparse-graph codes, and subsumes ANCC as a degenerated case. Further, while ANCC has a network codeword length in the order of $O(m)$, the effective network codeword length of GANCC is in the order of $O(Nm)$, where $m$ is the number of cooperating users and $N$ is the packet length for each user. Please note that the long effective code length of GANCC is achievable regardless of whether channel codes exist in each packet. Hence, GANCC requires significantly fewer users to cooperate than ANCC to attain a similar (network) coding gain.

Different code ensembles are investigated for use in GANCC, including *circulant low density generator matrix* (LDGM) codes and *circulant lower-triangular LDPC* (LT-LDPC) codes, both of which are natural extensions of LDGM and LT-LDPC codes used in ANCC, as well as the newly proposed *extended-circulant low-density generator-matrix* (EC-LDGM) codes. Additionally, two code constructing algorithms are discussed to optimize the actual code graph of these codes using locally-available information: The *column weight concentration* (CWC) algorithm provides a simple method to construct codes with balanced protection; and the *distributed progressive edge growth* (DPEG) algorithm improves the girth, and hence the code performance, at the cost of a larger complexity. The efficiency of GANCC using these proposed codes are further analyzed using density evolution. Finally, the realistic code performance is verified and benchmarked by computer simulations.

The remaining of this paper is organized as follows. Section II introduces the system model. Section III briefly introduces ANCC, and Section IV details the key idea and the general framework of GANCC. Section V discusses different code designs and code ensembles. Section VI conducts theoretical analysis. Finally, concluding remarks are provided in Section VIII.

## II. System model

The model of interest here comprises $m$ terminals communicating wirelessly to a common destination with only one antenna in each terminal by binary phase-shift keying (BPSK) modulation. In the first phase (broadcast phase), each terminal takes turns to broadcast its data packet, thereafter referred t o as *source-packet*. In the second phase (relay phase), each terminal takes turns to help forward others' data, thereafter referred to as *relay-packet* or *parity-packet*. The destination listens through both transmit phases, and combines all its reception to recover the source-packets for all the terminals.

We assume that all the communication channels used in this paper are spatially independent with fading coefficient $\alpha$ and channel noise $Z$. The fading coefficient $\alpha$ is modeled as a zero-mean, independent, circularly symmetric complex Gaussian random variable with unit variance, whose magnitudes $|\alpha|$ is Rayleigh distributed. We assume that the fading coefficient $\alpha$ is always known to the receivers but not to the transmitters. The channel noise $Z$ captures the addictive channel noise and interference, and is modeled as a complex Gaussian random variable with zero mean and variance $N_0$. For completeness, we consider both block fading (very slow fading) and independent and identically distributed (IID) fading (very fast fading). In the block fading scenario, the fading coefficient $\alpha$ remains constant during one round of user cooperation, and changes independently from one round to another. In the IID fading scenario, the fading coefficient changes independently from bit to bit.

## III. Overview of Adaptive Network Coded Cooperation

The ANCC protocol proposed in [8][9] operates as follows.

Consider an $m$-to-one data-collection network. In the first phase, each of the $m$ terminals airs a source-packet of length $N$ in its designated time slot. A terminal that is not transmitting listens, decodes what it hears, and collects the successfully decoded packets in its *retrieval-set*. Due to channel fading and other impairments, a terminal may not be able to retrieve all the packets.

In the second phase, each terminal randomly selects a small number of packets from its retrieval-set, computes their check-sum (i.e. XOR these binary vectors bit-by-bit) to form a length-$N$ relay-packet, and forwards it to the destination. Thus, by the end of the second phase, the $m$ terminals have transmitted, through user cooperation, a $(2m, m)$ network code in the form of a distributed, random, systematic LDPC code [8][9]. The source-packets transmitted in the first phase form the systematic symbols of the network code, and the relay-packets transmitted in the second phase constitute the parity symbols.

To help illustrate, consider a simple example of $m = 5$ users. Assume that for a particular round of cooperation, the *inter-user channels* form an instantaneous network topology as shown in Fig. 1, where a directed link in the figure represents a quality connection that will hold, say, until at least the end of this cooperation round. For simplicity, the destination is not shown in Fig. 1. Let the source-packet and relay-packet transmitted by user $j$ be indexed with $j$ and $m+j$, respectively. Thus, the retrieval-set of each user contains, respectively,

$$\mathcal{R}(1) = \{\mathbf{1}, \mathbf{4}, \mathbf{5}\},$$

$$\mathcal{R}(2) = \{1, \mathbf{2}, \mathbf{3}, \mathbf{5}, \mathbf{6}\},$$

$$\mathcal{R}(3) = \{\mathbf{1}, 2, \mathbf{3}, \mathbf{5}, 6, \mathbf{7}\},$$

$$\mathcal{R}(4) = \{\mathbf{1}, \mathbf{2}, \mathbf{4}, \mathbf{6}, 7\},$$

$$\mathcal{R}(5) = \{\mathbf{2}, \mathbf{3}, 4, 5, \mathbf{7}, \mathbf{8}, \mathbf{9}\}.$$

Suppose that the packets marked in bold font are selected (randomly) by each terminal to compute check sums. This results in an LT-LDPC network code with parity check matrix:

$$
H_{ancc} = 
\begin{array}{cccccccccc}
\phantom{.}1 & 2 & 3 & 4 & 5 & 6 & 7 & 8 & 9 & 10 \\
\end{array}
\begin{bmatrix}
\mathbf{1} & 0 & 0 & \mathbf{1} & \mathbf{1} & 1 & 0 & 0 & 0 & 0 \\
0 & \mathbf{1} & \mathbf{1} & 0 & \mathbf{1} & 1 & 1 & 0 & 0 & 0 \\
\mathbf{1} & 0 & \mathbf{1} & 0 & \mathbf{1} & 0 & 1 & 1 & 0 & 0 \\
\mathbf{1} & \mathbf{1} & 0 & \mathbf{1} & 0 & 1 & 0 & 0 & 1 & 0 \\
0 & \mathbf{1} & \mathbf{1} & \mathbf{1} & 0 & 0 & 1 & 1 & 1 & 1 \\
\end{bmatrix}
\tag{1}
$$

$$\underbrace{\phantom{xxxxxxxxx}}_{\text{systematic symbols}} \underbrace{\phantom{xxxxxxxxx}}_{\text{parity symbols}}$$

Due to the random and on-the-fly construction of the code, a small bit-map field needs to be included in each relay-packet, so that the destination knows how the checks are formed and can correspondingly replicate the code graph and perform message passing decoding. Since a different network code is constructed and transmitted with each new round of cooperation, the overall system performance represents that of the ensemble average rather than of any individual code. Further, such a topology requires the destination be equipped with an adaptive decoder architecture, which can be implemented, for example, using software-defined radio (SDR).

The network code in (1) takes the form of LT-LDPC codes. This is because users take turns (time-division) to transmit, such that one can continue to listen to relay-packets until its turn. In the case when users transmit simultaneously through frequency-devision or code-devision, each user will collect only source-packets, and the resulting network code will become an LDGM code, whose right part of the parity check matrix is an identity matrix instead of a lower-triangular matrix. Since weight-1 columns degrade the performance, LDGM codes in general exhibit a higher error floor and a slightly worse water-fall region than LT-LDPC codes [8][9].

Finally, depending on the quality of the user-destination channels or the residual power supply, a terminal may choose to relay a single time or multiple times, each time using a different relay-packet, or not to relay at all. This means that the network codes here need not be fixed-length or fixed-rate. Further, if there exists a simple feedback mechanism from the destination, then the users can keep generating and transmitting relay-packets, until the destination stops it. The resulting network code thus becomes a rateless sparse-graph code.

## IV. Generalized Adaptive Network Coded Cooperation (GANCC)

### A. An Illustrating Example

The ANCC protocol does not consider or exploit the channel code which may well exist in each source-packet. Since the network code length is solely dependent on the number of users $m$, it takes a large number of users to cooperate in order to achieve a good network coding gain. The associated delay and management overhead can be costly. Further, in the case when a large cluster of co-located users are not possible (e.g. in a mobile network or a small-scale network), the network code length may not be long enough to provide a desirable coding gain.

The proposed GANCC protocol provides a remedy to this problem by integrating the channel codes and the network code in one single codeword, making the effective code length from $2m$ to $2mN$, where $N$ is the length of each packet. The beauty of GANCC is that the channel codes now constitute an *integral* part of the network code, rather than being loosely connected to the network code via serial concatenation.

To best illustrate this, consider an extreme case where each source-packet contains only $N$ *uncoded* bits with no explicit channel coding. For simplicity, we consider the same 5-user example discussed in the previous section. The LDPC network code of ANCC, whose parity check matrix $H_{ANCC}$ is given in (1), is rather weak due to the short block size (and the existence of length-4 cycles). The lack of channel coding in each packet further eliminates the possibility to iteratively decode the network code and the channel code to improve performance.

Now GANCC remarkably changes the situation by a simple operation of *interleaving*. For each terminal, after selecting the packets from its retrieval-set, instead of computing their check-sums bit-by-bit in their original bit orders, it will interleave these length-$N$ bit-steams, each using a different scrambling pattern, before adding them together to compute parities. The resulting parity check matrix of this *joint network and channel code*, $H_{GANCC}$, is illustrated in Fig.3, where $\pi_{i,j}$ is a permutation of an identity matrix, whose row permutation pattern determines how user $i$ scrambles user $j$'s bit-stream. Mathematically, $H_{GANCC}$ is constructed by substituting each entry of $H_{ANCC}$ in (1) with an $N \times N$ square matrix, where "0"s are replaced by null matrices, and "1"s are replaced by independent permutation matrices, except for the "1"s on the right diagonal which are replaced by identity matrices (trivial permutations). In the degenerated case where all the permutation matrices use the identity matrix, then GANCC reduces to ANCC.

These permutation matrices or interleavers are critical to the performance of GANCC. First, interleaving integrates the bit-streams of all the users in one big network code, inter-connecting previously-unrelated bits for them to provide inference about one another. Since the effective code length becomes $O(mN)$, where $N$ typically ranges from a few hundred to a few thousand in practical systems, the system therefore obviates the need for many terminals to cooperate, making GANCC more practical. Second, by permuting each bit-stream using a different pattern, and so breaking the length-4 cycles that may previously exist in $H_{ANCC}$, interleaving reduces the chance for short cycles. In the example, $H_{ANCC}$ in (1) consists of several length-4 cycles, but the corresponding $H_{GANCC}$ in Fig. 3 has a much smaller fraction of length-4 cycles if any.

In GANCC, each terminal $i$ needs to generate or store a set of (random) interleaver $\pi_{i,j}$, $j = 1, 2, ..., m$, whose knowledge must be revealed to the common destination. This consumes a large storage space for all parties involved and/or a good amount of signaling overhead. To alleviate this burden, algebraic interleavers can be used in lieu of random interleavers [13]. An algebraic interleaver is one whose scrambling pattern can be generated on-the-fly using an often-recursive formula with a couple seeding parameters. Through the proper choice of formula and parameters, an algebraic interleaver can be made to behave much like a random interleaver, but requires significantly less storage [13].

In the context of GANCC, solutions that are even simpler than algebraic interleavers are possible by exploiting quasi-cyclic LDPC codes, or, circulant LDPC codes [14]. Several studies have shown that circulant matrices/interleavers can be used to construct good LDPC codes with simple encoding/decoding implementations. Thus, instead of using random or algebraic permutation matrices, we replace the "1" entries in $H_{ANCC}$ with $N \times N$ circulant matrices, such as the one shown below ($N = 4$):

$$\pi_{i,j} = \begin{bmatrix} 0 & 1 & 0 & 0 \\ 0 & 0 & 1 & 0 \\ 0 & 0 & 0 & 1 \\ 1 & 0 & 0 & 0 \end{bmatrix}. \tag{2}$$

Since each row is the right cyclic shift of the previous row, it takes a single parameter, the position of the non-zero entry in the first row, termed the offset and denoted as $p$, to determine a circulant matrix. In practice, it is possible to make $p$ a function of the terminals' indexes, thus eliminating any storage space and signaling overhead. The resultant code $H_{GANCC}$ is, by convention, denoted as an $(N, m, 2m)$ QC-LDPC code, where $N$ is the size of the circulant submatrices, and $m$ and $2m$ are the number of submatrices per column and per row respectively in the base matrix.

GANCC requires very little additional complexity than ANCC, but produces a random LDPC code whose code length is several magnitudes larger. Notice that $H_{GANCC}$ in general has a lower density than $H_{ANCC}$. In the example, $H_{ANCC}$ in (1) is rather dense, whereas its counterpart $H_{GANCC}$ in Fig. 2(A) appears to have just the right density. In practice, a delicate balance needs to be accounted for when choosing the check degrees, since heavy density breaks the message-passing decoding and excessive sparsity leads to uselessly weak codes.

To demonstrate the advantage of GANCC over ANCC and the efficiency of circulant inter-leavers, we simulate the 5-user example discussed previously, where the users each transmit an uncoded data-packet of length $N = 1000$ in the first phase and relay a parity-packet of length $N = 1000$ in the second phase. We evaluate three cases: ANCC with (10,5) LT-LDPC codes, and GANCC with (10000,5000) LT-LDPC codes based on random permutation matrices and circulant matrices, respectively. We plot both the bit error rate (BER), averaged over all the bits in all the data-packets, and the packet error rate (PER), averaged over all the users, versus the user-destination signal-to-noise ratio (SNR) in Fig.4. Since the network topology changes from time to time, resulting a different network graph and hence a different network code every time, the curves represent the ensemble *average* performance rather than that of a single code. We observe that random permutation matrices and circulant matrices make no performance difference in GANCC (curves overlapping), and they both significantly outperform ANCC, by 7 dB in BER and by 11 dB in PER. For fairness, we have considered block Rayleigh fading channels, such that there are only $m = 5$ different channel realizations in each codeword regardless of the code length. Hence, the impressive gains achieved by GANCC is only due to interleaving and the larger code length and the richer decoding context that come after. Foreseeably, when the channels become fast fading, large code lengths in GANCC will bring additional time diversity and consequently even larger performance advantage over ANCC.

## B. The General Framework

In general, the source-packet from each terminal is channel coded. Assume all the packets consume the same bandwidth $N$ (bits). Let $H_1, H_2, \cdots, H_m$ be the parity check matrices of the channel codes, $(N, K_1), (N, K_2), \cdots, (N, K_m)$, used in each source-packet, respectively, where $K_i$ is the raw data size for user $i$.

Irrespective of whether or not the source-packets are channel coded, each relay performs the same procedure as discussed before: after collecting a retrieval-set, (randomly) selects a few packets from the retrieval-set, interleaves them using different patterns for each, computes the parity-stream for these interleaved bit-streams, and forwards it to the destination.

Viewed from the destination, the combination of all the source-packets and the relay-packets together form one big network code whose parity check matrix consists of $2mN$ columns, corresponding to $\sum_i K_i$ raw data bits, and $2mN - \sum_i K_i$ rows, corresponding to $\sum_i (N - K_i)$ "channel-checks" and $mN$ "network-checks". Consider the 5-user example, the parity check matrix of the *unified network-channel code*, $H_{GANCC}$, will take the general form in Fig. 3, where $\pi_{i,j}$ is the (circulant) permutation matrix for user $i$ to interleaver user $j$'s data, and $H_i$ is the parity check matrix of the channel code used in user $i$'s source-packet.

The unified channel-network coding model depicted in Fig. 3 is general. It holds regardless of whether none, some, or all source-packets are channel coded and by what channel codes. When user $i$ does not employ a channel code, $H_i$ reduces to an identity matrix, which is like non-existent from the encoding and decoding perspective.

Three decoding strategies are available for GANCC. The optimal decoder treats $H_{GANCC}$ as one integrative code and performs joint channel-network decoding. This becomes practical when all the channel codes involved can be individually decoded by the message-passing algorithm, and so will the entire channel-network code. Alternatively, a two-level decoding architecture can be employed, where the network code, specified by the lower $mN$ rows of $H_{GANCC}$ in Fig. 3, is first decoded using the message-passing algorithm, whose soft (probabilistic) outcomes are subsequently passed to the individual channel codes for channel decoding. If complexity permits and if the channel codes produce soft outputs, these soft outputs may iterate back to the network code for successive refinement, enabling an iterative network-channel decoding architecture.

Note that sequential and iterative network-channel decoding also apply to ANCC, but a truly joint network-channel decoding on one unified code graph is possible only in GANCC, where the channel codes constitute an integral part of the network code through (circulant) interleaving. The example of uncoded source-packets in the previous section already revealed the critical importance of integrative coding. The coded example below further confirms this point. We use the same 5-user scenario over block Rayleigh fading channels, but the source-packets are now encoded by (2000,1000) (3,6)-regular LDPC codes. The simulation results, with solid lines for PER and dashed lines for BER, are shown in Fig. 5. "ANCC" and "GANCC1" are decoded using the same sequential decoding with 30 iterations of the base LDPC network code, followed by 30 iterations of the LDPC channel code; and the 3-4 dB gain of the latter is solely due to the circulant interleaving in GANCC. "GANCC2" and "GANCC3" improve "GANCC1" by feeding outputs from the channel codes back to the network code 5 and 10 times (5 and 10 global iterations), respectively. "GANCC4" performs joint network-channel decoding, whose efficiency is evident from the fact that it uses the same low complexity as "ANCC" and "GANCC1" (30 message-passing iterations on all checks and bits), but provides a performance on par with "GANCC3" (300 message-passing iterations on all checks and bits)!

## V. DISTRIBUTED CONSTRUCTION OF GOOD CODES

Randomly-constructed codes are easy to realize but do not always perform well. Below we discuss a few practical approaches to construct good codes. Putting aside the channel codes that may be present in some or all of the source-packets, the network code in GANCC is a circulant sparse-graph code, specified by a base LDPC code of, say, $(2m, m)$, and a set of $N \times N$ circulant matrices that replacing the "1" entries in the base LDPC code. Although the row-column connections are constrained, carefully-designed QC-LDPC codes perform on par with random LDPC codes especially at short length. The primary design challenge here is that codes must be constructed *distributedly* using only the incomplete information available locally. Below we start with circulant submatrices, and then move on to the design of the base LDPC code. Finally we propose a new class of circulant LDPC codes which can be exploited in GANCC and which may outperform circulant LT-LDPC and circulant LDGM codes under some circumstances.

### A. Circulant Matrices

Let $p_{i,j}$ be the offset of the circulant matrix that terminal $i$ uses to scramble terminal $j$'s data. Well-chosen $p_{i,j}$'s are not only storage-efficient, but also ensure a girth of at least 6 for the resultant circulant LDPC code.

**Theorem:** An $(N, s, t)$ QC-LDPC code has a girth $\geq 6$ if and only if for any two row indexes $0 \leq i_1 < i_2 < s$ and any two column indexes $0 \leq j_1 < j_2 < t$ in the base matrix, $p_{i_1,j_2} - p_{i_1,j_1} \neq p_{i_2,j_2} - p_{i_2,j_1} \mod \text{N}$.

**Remark:** The proof is not difficult and therefore omitted. The offset values of the circulant submatrices need not be different to rid of length-4 cycles. Rather, it is the separation between the offset values that matters. In a conventional QC-LDPC code, every entry in the base code is substituted with a circulant submatrix. In the proposed GANCC protocol, because of channel outage and deliberate de-selection of certain packets at each relay, some entries will be replaced by zero submatrices. This results in sparser QC-LDPC codes, whose matrix sub-divisions are larger than column/row weights and which generally outperform their denser counter-parts [16].

In GANCC, the submatrix size $N$ far exceeds the number of sub-division $m$, making it easy to select good $p_{i,j}$'s which are storage-efficient and achieve girth $\geq 6$ at the same time. One possibility, for example, is $p_{i,j} = ij$, for $0 \leq i \leq m - 1$ and $0 \leq j \leq 2m - 1$.

### B. The Base LDPC Code

Now consider the base code in GANCC. Irregular LDPC codes can outperform regular ones, but the degree profile must be designed to match the physical channel, which, in the case of GANCC, is an $m$-cyclic channel comprising $m$ independently faded segments. It is not only difficult to optimize the degree profile for such a channel, but the lack of central control also makes it impossible to construct a code with the the target column and row degree profile.

An LDGM or LT-LDPC code is regular if the systematic part of its parity check matrix has uniform column weight (thereafter referred to as the *degree*) and near-uniform row weight. For a regular LT-LDPC code, the column weight of the lower-triangular part should also decrease proportionally to preserve a uniform density. Our design below focuses on the *edge connection* of each individual code (rather than the degree profile of the code ensemble).

**Design I: Column Weight Concentration Algorithm**

When each relay randomly selects packets from its retrieval-set, it is highly likely that source-packets get unequal protection, with some over-protected and others under-protected. This should be differentiated from irregular LDPC codes, whose degree profile is carefully designed to optimize a "wave" phenomenon. Here the insufficiently-protected source-packets are vulnerable to errors and will likely degrade the overall system performance.

The *column weight concentration algorithm* aims at making the column weights as uniform as possible. The idea is simple and easy to implement: when each user listens and decodes packets, it also keeps track of the number of checks each packet participates in; when its turn comes, it selects from its retrieval-set the ones that are least protected to form a check. Due to possible inter-user outage, each user has only a partial view of the node degrees of the network code. The resulting code may not be exactly regular, but this simple mechanism effectively eliminate the majority of null-weight or weight-one columns in the $H$ matrix, which are most harmful to the code performance. We note that the network codes simulated in Figures 4 and 5 are constructed distributedly using this CWC algorithm. Otherwise, the performance of the individual network code will vary significantly (especially since the network size is small), with some instances yielding rather disappointing performance.

**Design II: Distributed Progressive Edge Growth Algorithm**

For a given degree profile, the progressive edge growth (PEG) algorithm [15] builds a Tanner graph by connecting variable nodes and check nodes edge by edge, such that each time the added edge has minimal impact on the girth of the graph. An effective tool to maximize the girth, the PEG algorithm has resulted in some of the best known codes at short lengths [15].

The PEG algorithm exploited to the base LDPC code here is somewhat different from that in the original proposal [15]. First, the algorithm will be run in a distributed rather than centralized manner. A terminal continues to hear and collect information on the growth of the graph structure until it has fulfilled its turn of message forwarding. Each terminal independently constructs a subgraph to the level of its interest based on the information available locally, to determine how to add edges (i.e. form the parity check). Due to possible channel outage, the graph envisioned by each terminal may be slightly different from the true code graph. Second, because of the lack of global knowledge and central control, the terminals have no knowledge of the resultant degree of any of the variable nodes. Hence the graphs can not expand from the variable nodes as described in [15], but will instead expand from the check nodes. The general idea remains similar, and the process is summarized below, which is dual to what is discussed in [15].

**Distributed Progressive Edge Growth (DPEG) Algorithm**;

$d_{c_j}$: degree of check node $c_j$;

$E_{c_j}^k$: $k$th edge incident to check node $c_j$;

$N_{c_j}^l$: set of variable nodes within depth $l$ from check node $c_j$ in the subgraph;

$\bar{N}_{c_j}$: set of variable node complementary to $N_{c_j}$;

$edge(v_i, c_j)$: edge connecting the $i$'th variable node and the $j$'th check node;

**for** $j = 0$ **to** $m - 1$ **do**

    **for** $k = 0$ **to** $d_{c_j} - 1$ **do**

        **if** $k = 0$ **then**

            $E_{c_j}^0 \leftarrow edge(v_i, c_j)$, where $v_i$ is a variable node having the lowest estimated variable degree under current graph setting $E_{c_0} \cup E_{c_1} \cup \cdots \cup E_{c_{j-1}}$;

        **else**

            Expand a tree from $c_j$ up to depth $l$ under the current graph setting such that $\bar{N}_{c_j}^l \neq \emptyset$ but $\bar{N}_{c_j}^{l+1} = \emptyset$, or the cardinality of $N_{c_j}^l$ stops increasing but is less than $m$, then $E_{c_j}^k \leftarrow edge(v_i, c_j)$, where $v_i$ is one variable node picked from the set $\bar{N}_{c_j}^l$ having the lowest estimated variable node degree;

        **end**

    **end**

**end**

*C. Extended Circulant LDGM Codes*

As discussed in Section III, in some scenarios, the resulting network code may take the form of LDGM codes. Due to the large number of weight-1 columns, whose outbound reliability information never gets updated, (circulant) LDGM codes produce a higher error floor as well as a worse water-fall region than LT-LDPC codes. To alleviate the negative impact of weight-1 columns, we propose the *extended circulant low-density generator-matrix (EC-LDGM) codes*.

The only difference between the new EC-LDGM code and the circulant LDGM code discussed before is an additional differential encoding process. Upon its turn to relay, a terminal first performs the same process as discussed before: select data-packets, circularly shift these bit-streams, and XOR them bit-by-bit to obtain the bit-stream of the parity-packet. Then, instead of sending this bit-stream $\{x_i\}$ as is, the relay sends the *differentially encoded* version $\{y_i\}$: $y_0 = x_0$, and $y_i = \text{xor}(x_i, x_{i-1})$, for $i = 1, 2, \cdots, N-1$. What this reflects in the parity check matrix, as illustrated in Fig. 2(C), is that the right diagonal blocks now have zigzag pattens instead of a single line in the main diagonal.

Clearly, EC-LDGM codes drastically reduce the number of weight-1 columns from $mN$ to $m$, which in term improves the performance over conventional circulant LDGM codes. Comparing to circulant LT-LDPC codes, EC-LDGM codes have an advantage on IID fading channels, but fall short on block fading channels. This is because, on block fading channels, all the $N$ bits in the same circulant block are faded together. Most checks in an EC-LDGM codes involve two (parity) bits coming from the same block, which severely undermines the usability of these checks when these bits experience the same deep fade. In contrast, all the checks in a circulant LT-LDPC code consist of bits from different blocks, and therefore provide a much better diversity. On the other hand, an EC-LDGM code has only $m$ weight-1 columns (one in each parity block, see Fig. 2(C)), whereas a circulant LT-LDPC code has $N(>> m)$ weight-1 columns (the entire last parity block, see Fig. 2(A)). Hence, on IID fading channels where all the bits experience different fades, EC-LDGM codes can outperform circulant LT-LDPC codes.

## VI. Performance Analysis

To predict the performance of GANCC using LDGM, LT-LDPC and EC-LDGM codes, we conduct theoretic analysis using density evolution [18]. To make analysis tractable and reflective (primarily) of the role of network coding, we assume source-packets have no channel coding.

### A. The General Formulation

Consider a block fading scenario. Let the number of spatially independent channels $m$ fixed to some finite value, and let the size of each uncoded data-packet $N$ increase without bound. The overall network code has length $2Nm \to \infty$. Each codeword is transmitted over $m$ independent Gaussian channel realizations with noise variances $\sigma^2(t)$, $t = 1, 2, \cdots, m$.

In density evolution, it is an accepted practice to assume that the log-likelihood ratios (LLR) extracted from individual Gaussian channels as well as those exchanged between different types of variable and check nodes all follow some Gaussian density. This Gaussian approximation was initially proposed as a convenient tool for iterative analysis of LDPC codes, but one that is largely pragmatic [19]. Recent research has provided solid statistical support to the accuracy of this assumption [17]. Combined with a symmetry condition, the Gaussian approximation leads to a particularly useful result that the variance of the Gaussian LLRs equals twice the mean value [19]. Hence, the evolution of the LLR messages and the corresponding error probabilities can be characterized by a single parameter, the mean of the LLR messages.

Let $u_0(q)$, $q = 1, 2, \cdots, m$, be the LLR messages extracted from the $q$th Gaussian channel, where $u_0(q)) \sim \mathcal{N}(\mu_{u0}(q), 2\mu_{u0}(q))$ and $\mu_{u0}(q) = 2/\sigma^2(q)$. Let $u_0$ be the average of $u_0(q)$s, averaged over the distribution of edges in the code graph that are associated with $u_0(q)$. Let $u^{(l)}$ and $v^{(l)}$ be the LLR messages passed from variable nodes to check nodes, and from check nodes to variable nodes, in the $(l)$th iteration, respectively. Further, let $\mu_{u0}$, $\mu_u^{(l)}$ and $\mu_v^{(l)}$ be the respective means of $u_0$, $u^{(l)}$ and $v^{(l)}$. Since the codeword experiences different Gaussian channel realizations and since the network code is an irregular sparse-graph code, the messages generated at each one or half decoding iteration are in fact mixed Gaussian. Following the convention of density evolution, we assume that the messages passed across the edges from one type of nodes to the other are independent and follow the same Gaussian distribution.

Let $d_v$ and $d_c$ be the maximum degree of variable nodes and check nodes. Let $\lambda(x) = \sum_{i=1}^{d_v} \lambda_i x^{i-1}$ and $\rho(x) = \sum_{j=1}^{d_c} \rho_j x^{j-1}$ be the edge-perspective degree profile of the variable nodes and check nodes, respectively, where $\lambda_i$ and $\rho_j$ are the percentages of edges connecting to variable nodes of degree $i$ and check nodes of degree $j$, respectively. Following the standard density evolution procedure of irregular sparse-graph codes [19], we get

$$\mu_u^{(l)} \approx \sum_{j=2}^{d_c} \rho_j \Psi^{-1}\left(\left[\sum_{i=1}^{dv} \lambda_i \Psi\left(\mu_{u0} + (i-1)\mu_u^{(l-1)}\right)\right]^{j-1}\right), \tag{3}$$

where

$$\mu_u^{(0)} = 0, \qquad \Psi(\mu_x) \triangleq E\left[\tanh\frac{x}{2}\right] = \frac{1}{\sqrt{4\pi\mu_x}} \int_{-\infty}^{\infty} \tanh\frac{x}{2} e^{-\frac{(x-\mu_x)^2}{4\mu_x}} \, dx, \tag{4}$$

where $x$ is a Gaussian variable with mean $\mu_x$ and variance $2\mu_x$. To simplify computation, one may use approximations $\Psi(\mu_x) \approx 1 - \exp(-0.4527\mu_x^{0.86} + 0.0218)$ [19] or $\Psi(\mu_x) \approx 1 - \exp(-0.432\mu_x^{0.88})$ [17], where the former works for $0 < \mu_x \leq 10$, and the latter works for the entire region of $\mu_x > 0$.

At the end of the $l$th iteration, the total LLR messages associated with the variable nodes having degree $i$ and transmitted through the $q$th channel have a mean value of:

$$\mu^{(l)}(i, q) = \mu_{u0}(q) + i \cdot \mu_u^{(l-1)}. \tag{5}$$

The error probability associated with degree-$i$ variable nodes on the $q$th channel is: $p_e^*(i, q) \approx Q\left(\sqrt{\mu^{(l)}(i, q)/2}\right)$. Averaging over all the $m$ channel realizations and all the variable nodes, we get the error probability of this LDPC code ensemble

$$p_e^* \approx \sum_{i=2}^{dv} \xi_i' \sum_{q=1}^{m} \frac{1}{m} p_e^*(i, q) = \sum_{i=2}^{dv} \xi_i' \sum_{q=1}^{m} \frac{1}{m} Q\left(\sqrt{0.5\mu^{(l)}(i, q)}\right), \tag{6}$$

where $\xi_i'$ is the percentage of systematic variable nodes (from the node perspective) having degree $i$, and $\sum i = 2^{dv}\xi_i' = 1$.

Finally, the average error probability on block fading channels is the expectation over all the co-operation rounds, each round associated with a set of channel realizations $(\sigma^2(1), \sigma^2(2), ..., \sigma^2(m))$ drawn from their respective Rayleigh fading distributions: bellow:

$$p_e = \mathbf{E}_{\sigma^2(1), \sigma^2(2), ..., \sigma^2(m)}[p_e^*]. \tag{7}$$

**Remark:** The error rate formulated here estimates the ensemble-average performance of $(2mN, mN)$ LDPC codes ($N \to \infty$) over the so-called $m$-cyclic block fading channels. The joint network-channel codes we designed for GANCC are circulant LDPC codes, which constitute a subset of the general code ensemble. Here we take the results for the general ensemble to approximate that of the circulant sub ensemble. Although not entirely accurate, the difference will be very small, since (i) circulant LDPC codes are shown to perform on par with a randomly-generated LDPC code of the same degree profile, and (ii) at very large block sizes, a concentration rule holds such that almost all the codes in the ensemble perform very close to the ensemble-average performance.

We now evaluate the three classes of circulant codes proposed for GANCC: LDGM codes, LT-LDPC codes, and EC-LDGM codes.

### B. LDGM Codes

Circulant LDGM codes, shown in Fig. 2(B), are a special class of irregular LDPC codes. whose analysis fits in the general model discussed before. Consider the ensemble of rate-1/2 degree-$D$ LDGM codes. The check nodes have uniform degree $D + 1$, and the variable nodes have degrees $D$ and $1$, corresponding to the systematic bits and the parity check bits, respectively. Hence the variable and check degree profile is:

$$\lambda(x) = \sum_{i=1}^{d_v} \lambda_i x^{i-1} = \frac{1}{1+D}x^0 + \frac{D}{1+D}x^{D-1}, \tag{8}$$

$$\rho(x) = \sum_{j=2}^{d_c} \lambda_j x^{j-1} = x^D. \tag{9}$$

Since all the systematic variable nodes have degree $D$, the degree distribution $\xi'(x)$ becomes:

$$\xi'(x) = \sum_{i=1}^{d_v} \xi_i' x^{i-1} = x^{D-1}. \tag{10}$$

Gathering (3)-(7), and inserting the degree profiles (8), (9) and (10), we get the ensemble-average BER of the LDGM codes.

### C. LT-LDPC Codes

Following the same argument, (circulant) LT-LDPC codes, shown in Fig. 2(A), can be evaluated by inserting the right degree profiles to (3)-(7). For rate-1/2 circulant LT-LDPC codes with degree-$D$ and a balanced density in the parity check matrix, we will have (ideally)

$$\lambda(x) = \sum_{i=1}^{D-1} \frac{2i}{D(3D+1)} x^{i-1} + \frac{2(D+1)}{3D+1} x^{D-1}, \tag{11}$$

$$\rho(x) = \sum_{j=D+1}^{2D} \frac{2(j-D)}{(D+1)(3D+1)} x^{j-1}, \tag{12}$$

$$\xi'(x) = x^{D-1}. \tag{13}$$

### D. EC-LDGM Codes

In circulant LDGM and LT-LDPC codes, each $N \times N$ block has only one weight per row, so the bits participating in any one check experience different channel fades. In an EC-LDGM code, as shown in Fig. 2(C), the two parity bits coming from the same block experiences the same channel fade[1], and the message exchange between these parity bits is confined to the same block. Thus, one needs to track the LLR density of the $mN$ systematic variable nodes (collectively, denoted as $\mathfrak{R}_s$), and the $mN$ parity variable nodes of the network code separately.

Consider the parity checks contributed by the $t$th user. Let $\mu_{us}(t)$ and $\mu_{up}(t)$ be the respective mean LLR values from the systematic variable nodes and the parity variable nodes to the check nodes, at the $t$th user. Let $\mu_{vs}(t)$ and $\mu_{vp}(t)$ be the respective mean LLR values from the check nodes to the systematic and the parity variable nodes at the $t$th user. We have

$$\Psi(\mu_{us}(t)) = \mathbf{E} \left[ \prod_{p \in \mathfrak{R}_s \setminus t} \Psi(\mu_{vs}(t)) \right] \Psi^2(\mu_{vp}(t)) \approx \left[ \sum_{j=1}^{m} \Psi(\mu_{vs}(j)) \right]^{|\mathfrak{R}_s|-1} \Psi^2(\mu_{vp}(t)), \tag{14}$$

$$\Psi(\mu_{up}(t)) = \mathbf{E} \left[ \prod_{p \in \mathfrak{R}_s} \Psi(\mu_{vs}(t)) \right] \Psi(\mu_{vp}(t)) \approx \left[ \sum_{j=1}^{m} \Psi(\mu_{vs}(j)) \right]^{|\mathfrak{R}_s|} \Psi(\mu_{vp}(t)). \tag{15}$$

As the decoding process proceeds from the $(l-1)$th iteration to the $l$th iteration, the LLR means evolve as follows:

$$\mu_{vs}^{(l)}(t) = \mu_{u0}(t) + (d_v - 1) \, \bar{\mu}_{us}^{(l-1)}, \tag{16}$$

$$\mu_{vp}^{(l)}(t) = \mu_{u0}(t) + \mu_{up}^{(l-1)}(t), \tag{17}$$

---

[1]There exist $1/N$ of checks that involve one parity bit each, but this percentage vanishes as $N \to \infty$.

where $\bar{\mu}_{us}^{(l)} = \frac{1}{m} \sum_{j=1}^{m} \mu_{us}^{(l)}(j)$ is the average LLR mean, averaged over all the checks from all the $m$ users. Gathering (14), (15), (16) and (17) yields

$$\mu_{us}^{(l)}(t) \approx \sum_{j=2}^{d_c} \rho_j \upsilon \Psi^{-1}\left( \left[ \sum_{i=2}^{dv} \lambda_s(i) \sum_{k=1}^{m} \Psi\left(\mu_{u0}(k) + (i-1)\bar{\mu}_{us}^{(l-1)}\right) \right]^{|\mathfrak{R}_s|-1} \Psi^2\left(\mu_{u0}(t) + \mu_{up}^{(l-1)}(t)\right) \right), \quad (18)$$

$$\mu_{up}^{(l)}(t) \approx \sum_{j=2}^{d_c} \rho_j \Psi^{-1}\left( \left[ \sum_{i=2}^{dv} \lambda_s(i) \sum_{k=1}^{m} \Psi\left(\mu_{u0}(k) + (i-1)\bar{\mu}_{us}^{(l-1)}\right) \right]^{|\mathfrak{R}_s|} \Psi\left(\mu_{u0}(t) + \mu_{up}^{(l-1)}(t)\right) \right), \quad (19)$$

where

$$\lambda_s(i) = \frac{\text{number of edges connecting to all systematic variable nodes of degree } i}{\text{number of edges connecting to all systematic variable nodes}}.$$

Finally, the estimation of the error probability at the end of the $l$th decoding iteration is computed similarly to that of circulant LDGM and circulant LT-LDPC codes:

$$p_e^* \approx \sum_{i=2}^{dv} \xi(i) \sum_{k=1}^{m} \frac{1}{m} Q\left(\sqrt{\mu_{vs}^{(l)}(i,k)/2}\right), \quad (20)$$

where $\mu_{vs}^{(l)}(i,k)$ is the mean LLR value computed by the degree-$i$ systematic variable nodes associated with the $t$th user.

## VII. Experimental Results

We conduct extensive simulations to verify the efficiency of the proposed GANCC framework and the different code constructing methods.

Fig. 6 compares the ensemble average performance of the circulant LT-LDPC network-channel codes constructed using the CWC algorithm and the DPEG algorithm. We evaluate $m = 5$ and $m = 10$ terminals on block Rayleigh fading channels. All the packets (symbols) have 1000 raw data bits without channel coding. Whereas the CWC algorithm is simpler, the DPEG algorithm performs better, providing about 1 dB additional gain at BER of $10^{-5}$ in both cases.

We also simulate different code ensembles, circulant LDGM codes, extended circulant LDPC codes, and circulant LT-LDPC codes, and compare them against the theoretical results obtained by the density evolution method. To get long codes, here we extend each source-packet to $N = 5000$ uncoded bits, but keep the number of users the same: $m = 5, 10$. The results, plotted in Fig. 7 and Fig. 8 respectively, show that the simulations match with the theoretically analysis fairly well with a gap of no more than $0.5$ dB between them. As expected, the LDGM ensemble performs the worst among the three code ensembles, EC-LDGM ensemble performs 2 dB better and the LT-LDPC ensemble performs an additional 1-2 dB better.

Whereas the slow fading case is where user cooperation becomes most useful, for completeness, we also examine the same simulation setup in an IID fading scenario. The performance of these codes constructed using CWC are plotted in Fig. 9. It should be noted that since the Gaussian approximation is less accurate on IID fading channels, the analytical results provided here become less accurate. This may explain why there is a relatively large gap of about 2-3 dB between the simulation and the analytical results; but it is assuring that they both exhibit the same qualitative trends. First, the circulant LDGM ensemble always performs worst due to the large number of harmful weight-1 columns in the code. Second, the circulant LT-LDPC ensemble has a rather obvious error floor in its BER curves, due to the many weight-1 columns in the last one or the last few circulant blocks in the parity check matrix. Third, the EC-LDGM ensemble shows only water-falls in the BER region of interest, and thus becomes the best-performing code ensemble in the IID fading scenario.

## VIII. CONCLUSIONS

We have investigated distributed and adaptive wireless user cooperation in a multiple-sender single-destination scenario. Unlike other approaches that use fixed network coding schemes and therefore rely on the ideal assumption of no link outage, the *adaptive network coded cooperation* protocol developed in [8] cleverly constructs adaptive sparse-graph network codes to match to the constantly-changing network topology. Since ANCC alone is best suited for large large networks, this paper extends ANNC to *generalized ANCC*, or, GANCC, by integrating adaptive network coding with channel coding in the framework of circulant LDPC codes. Through code design, theoretical analysis and computer simulations, we show that GANCC achieves impressive coding performance even when there are only a limited number of cooperating terminals.

## REFERENCES

[1] T. E. Hunter and A. Nosratinia, "Coded Cooperation under slow fading, fast fading and power control," *Proc. Asilomar Conf. Signals, Syst., Comput.,* 2002.

[2] M. Janani, A. Hedayat, T. Hunter, and A. Nosratinia, "Coded cooperation in wireless communications: space-time transmission and iterative decoding," *IEEE Trans. Sig. Processing,* pp. 362-371, Feb 2004.

[3] J. N. Laneman, G. W. Wornell, "Distributed Space-Time-Coded protocols for exploiting cooperative diversity in wireless networks," *IEEE Trans. Inform. Theory,* vol. 49, NO. 10, Oct 2003, pp. 2415-2425.

[4] G. Kramer, M. Gastpar, and P. Gupta, "Cooperative strategies and capacity theorems for relay networks," to appear *IEEE Trans. Inform. Theory,* 2006.

[5] R. Ahlswede, N. Cai, S.-Y.R. Li and R.W. Yeung, "Network information flow", IEEE-IT, vol. 46, pp. 1204-1216, 2000.

[6] R. Koetter, and M. Medard, "An algebraic approach to network coding," *IEEE Trans. Networking,* Oct. 2003.

[7] M. Effros, M. Médard, T. HO, S. Ray, D. karger, and R. Koetter, *"Linear network codes: a unified framework for source, channel and network coding,",* DIMACS workshop on network information theory, 2003.

[8] X. Bao, and J. Li, "Matching Code-on-Graph with Network-on-Graph: Adaptive Network Coding for Wireless Relay Networks," *Proc. Allerton Conf. on Commun., Control and Computing* IL, Sept. 2005.

[9] X. Bao, and J. Li, "Adaptive Network Coded Cooperation (ANCC) for Large Wireless Relay Networks," *IEEE Trans on Wireless communications,* vol. 7, no 2, pp. 574-583, Feb. 2008.

[10] X. Bao, and J. Li, "On the Outage Properties of Adaptive Network Coded Cooperation (ANCC) in Large Wireless Networks," *IEEE Intl Conf. on Acoustics, Speech, and Signal Processing* Toulouse, France, May. 2006.

[11] Y. Chen, S. Kishore, and J. Li, "Wireless diversity through network coding," *Proc.IEEE Wireless Commun. Networking Conf.,* March, 2006.

[12] C. Hausl, F. Schreckenbach and I. Oikonomidis, "Iterative network and channel decoding on a tanner graph", *Proc. Allerton Conf. on Commun., Control and Computing,* Urbana Champaign, IL, Sept. 2005.

[13] O. Y. Takeshita, and D. J. Costello, Jr. "New Classes Of Algebraic Interleavers for Turbo-Codes," *Proc. IEEE Intl Symp on Inform. Theory,* Boston, Aug. , 1998.

[14] Y. Kou, H. Tang, S. Lin, and K. Abdel-Ghaffar, "On Circulant Low Density Parity Check Codes," *Proc. IEEE Intl Symp. Inform. Theory,* p. 200, June 2002.

[15] X. Hu, E. Eleftheriou, and D. Arnold, "Regular and irregular progressive edge-growth tanner graphs", *IEEE Trans. on Inform. Theory,* Vol. 51, ISS. 1, pp. 386 - 398, Jan. 2005.

[16] Y. Mao, and A. H. Banihashemi, "A heuristic search for good low-density parity-check codes at short block lengths," *Proc. of IEEE Intl. Conf. Commun.,* pp 41-44, June 2001.

[17] K. Xie and J. Li, "On accuracy of Gaussian assumption in iterative analysis for LDPC codes," *Proc. Intl. Symp. Info. Theory,* Seattle, WA, June 2006.

[18] T. Richardson and R. Urbanke, "The capacity of low-density parity-check codes under message-passing decoding", *IEEE Trans. on inform. Theory,* Vol. 47, No. 2, Feb. 2001.

[19] S. Chung, T. Richardson, and R. Urbanke, "Analysis of sum-product decoding of low-density parity-check codes using a Gaussian approximation", Vol. 47, No. 2, Feb, 2001.

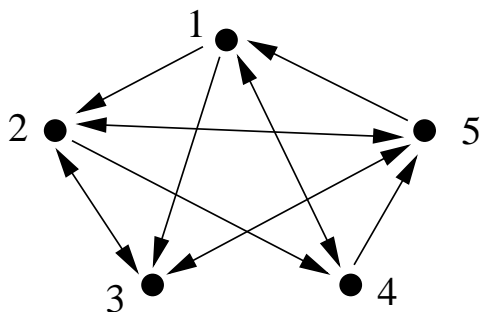

Fig. 1. An example of 5 users sending data to a common destination.

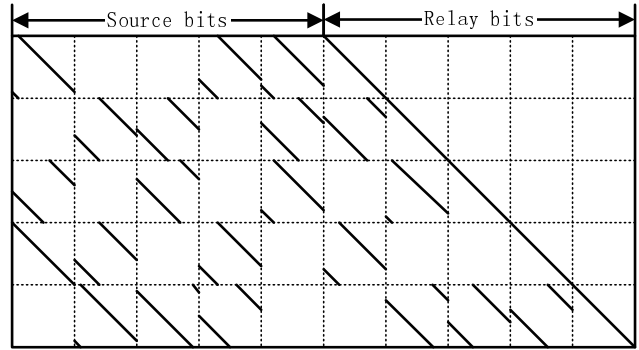

(A)

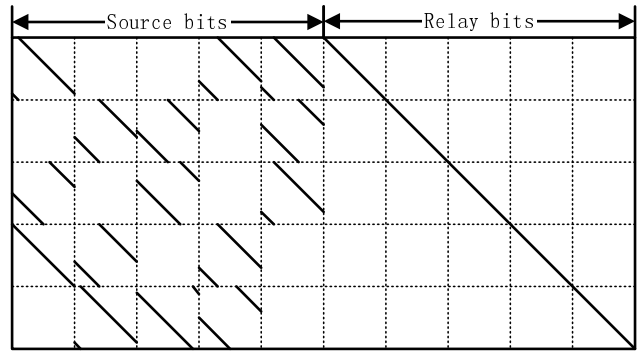

(B)

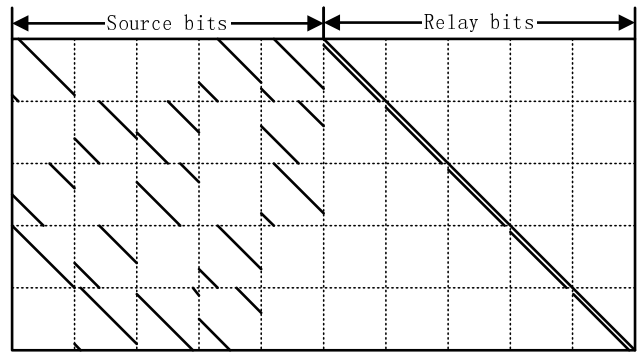

(C)

Fig. 2. Parity check matrix of (A) circulant LT-LDPC codes, (B) circulant LDGM codes, and (C) extended circulant LDGM codes.

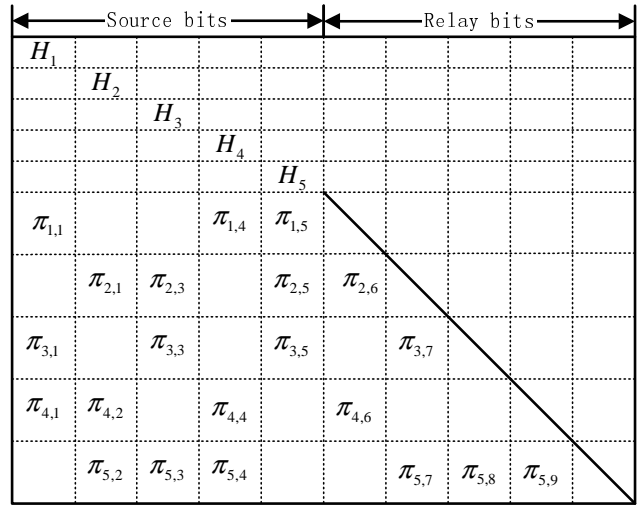

Fig. 3. An example of the $H$ matrix for the unified channel-network code used in GANCC with source-packets of H matrix $H_1, H_2, \cdots, H_5$ ($m = 5$ terminals).

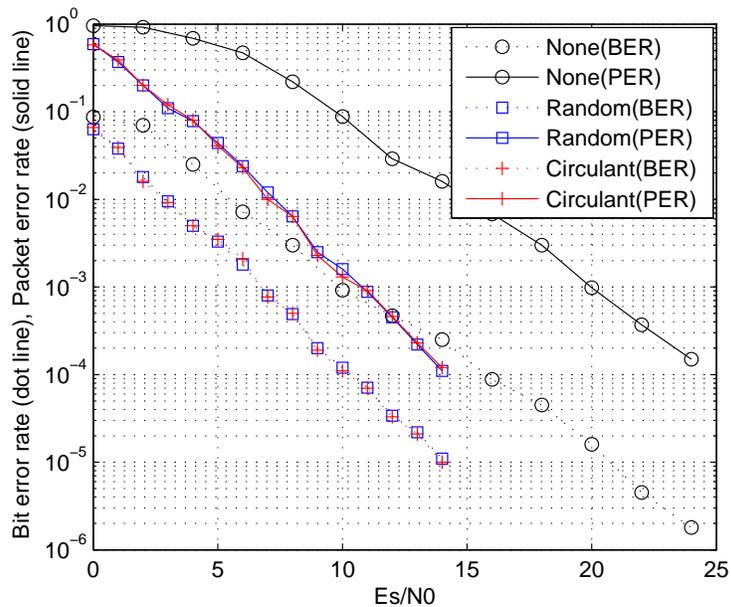

Fig. 4. The performance of GANCC with no explicit channel codes. $m = 5$, $N = 1000$.

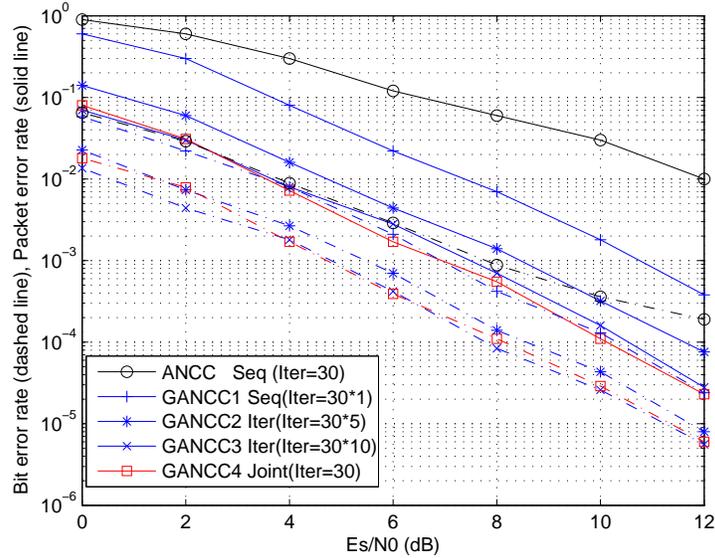

Fig. 5. Comparison between GANCC and ANCC with different decoding methods. $m = 5$, $N = 2000$. Each source-packet is channel coded with a $(3, 6)$-regular LDPC code, and circulant interleavers are adopted in GANCC. Solid lines: PER; dashed lines: BER.

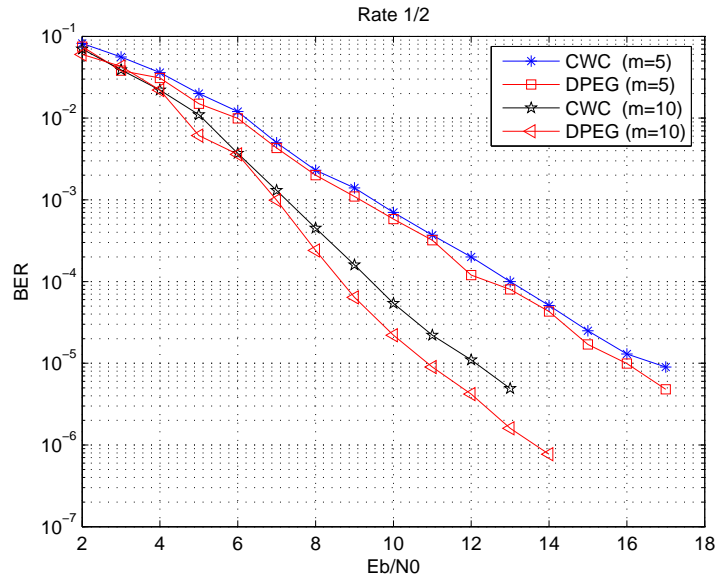

Fig. 6. Performance comparison between the *column weight concentrate* algorithm and the *distributed progressive edge growth* algorithm for LT-LDPC codes

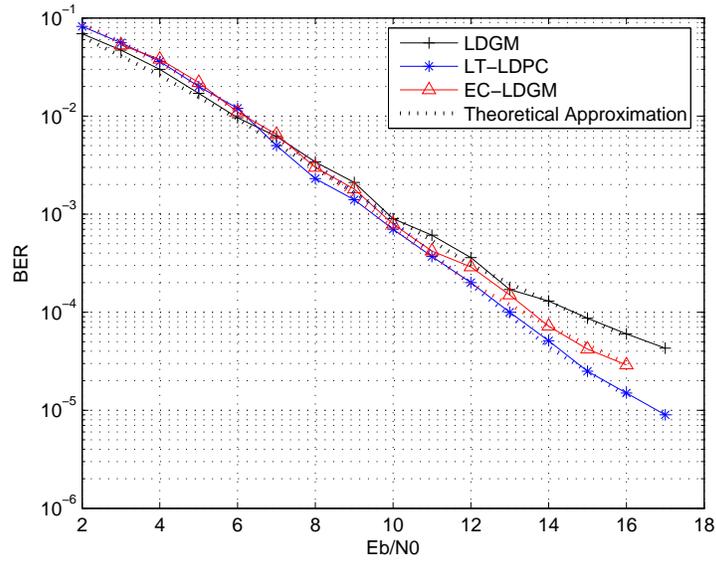

Fig. 7. Performance of different code ensembles with $m = 5$ terminals in block fading, code rate equals 1/2, and $N = 5000$.

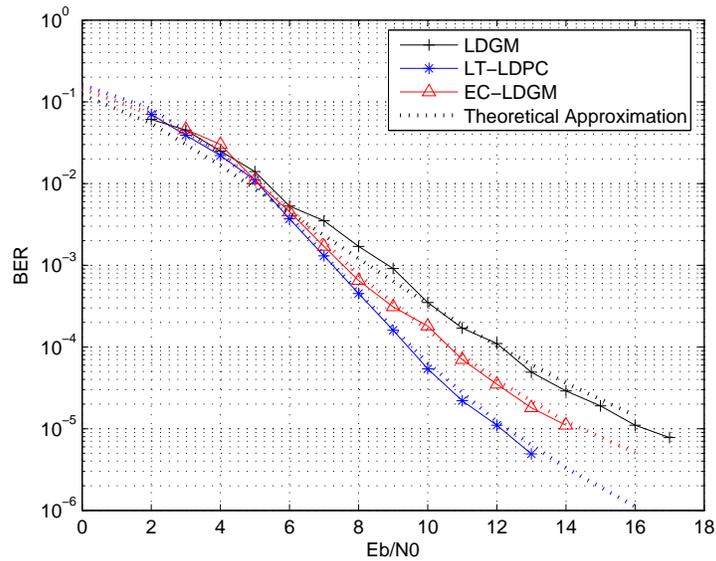

Fig. 8. Performance of different code ensembles with $m = 10$ terminals in block fading. Network code rate is 1/2, and $N = 5000$ uncoded bits.

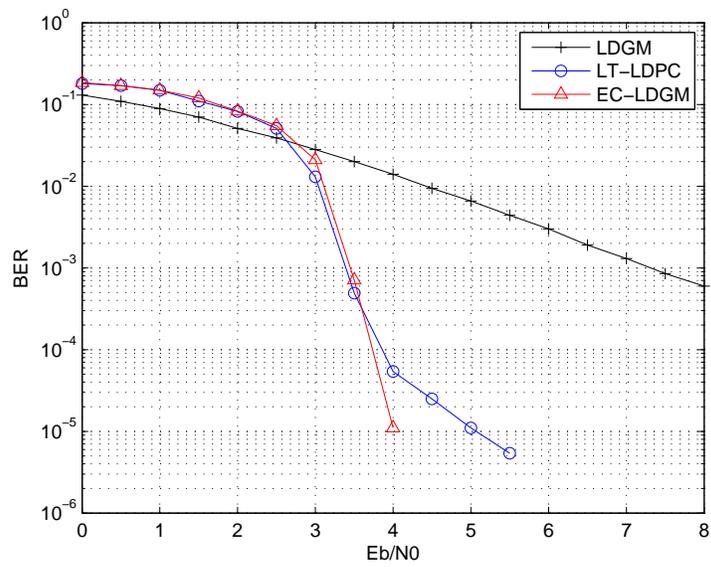

Fig. 9.   Performance of different code ensembles in IID fading. Network code rate is 1/2, and $N = 5000$ uncoded bits.